\documentclass[vecphys]{svmult}

\usepackage{amsmath}
\usepackage{makeidx}         
\usepackage{graphicx}        
\usepackage{multicol}        
\usepackage{cite}            
\usepackage[bottom]{footmisc}

\makeindex             

\newcommand{\atil}{\tilde{\alpha}}
\newcommand{\btil}{\tilde{\beta}}
\newcommand{\etil}{\tilde{\eta}}
\newcommand{\dtil}{\tilde{\delta}}
\newcommand{\Dtile}{\tilde{\Delta}_e}

\begin{document}

\title{Universality in modelling non-equilibrium pattern formation in polariton condensates}
\titlerunning{Universality in modelling pattern formation}
\author{N.G Berloff and J. Keeling}
\institute{Department of Applied Mathematics and Theoretical Physics, University of Cambridge, CB3 0WA, UK \texttt{N.G.Berloff@damtp.cam.ac.uk}
\and SUPA, School of Physics and Astronomy, University of St Andrews, KY16 9SS, UK
\texttt{jmjk@st-andrews.ac.uk}}

\maketitle

\begin{abstract}
   The key to understanding the universal behaviour of systems driven
  away from equilibrium lies in the common description obtained when
  particular microscopic models are reduced to order parameter
  equations.  Universal order parameter equations written for complex
  matter fields are widely used to describe systems as different as
  Bose-Einstein condensates of ultra cold atomic gases, thermal
  convection, nematic liquid crystals, lasers and other nonlinear
  systems.  Exciton-polariton condensates recently realised in
  semiconductor microcavities are pattern forming systems that lie
  somewhere between equilibrium Bose-Einstein condensates and lasers.
  Because of the imperfect confinement of the photon component,
  exciton-polaritons have a finite lifetime, and have to be
  continuously re-populated. As photon confinement improves, the
  system more closely approximates an equilibrium system.  In this
  chapter we review a number of universal equations which describe
  various regimes of the dynamics of exciton-polariton condensates:
  the Gross-Pitaevskii equation, which models weakly interacting
  equilibrium condensates, the complex Ginsburg-Landau equation---the
  universal equation that describes the behaviour of systems in the
  vicinity of a symmetry--breaking instability, and the complex
  Swift-Hohenberg equation that in comparison with the complex
  Ginsburg-Landau equation contains additional nonlocal terms
  responsible for spacial mode selection. All these equations can be
  derived asymptotically from a generic laser model given by
  Maxwell-Bloch equations. Such an universal framework allows the unified
  treatment of various systems and continuously cross from one system
  to another. We discuss the relevance of these equations, and their
  consequences for pattern formation.
\end{abstract}

\section{Introduction}

For a dissipative macroscopic system in thermal equilibrium,
relaxation toward a state of minimum free energy determines the states
that the system may adopt, and any possible pattern formation.  In
contrast, if a system is driven out of equilibrium by external fluxes,
then no such simple description is possible.  i.e., if a system may
exchange particles and energy with multiple baths (reservoirs), then
the states the system adopts depend not only on the temperatures and
chemical potentials of these reservoirs, but also on the rate at which
particles and energy are injected and lost from the system.  This
can not generally be captured by relaxation to minimise a given energy
functional

Both equilibrium and non-equilibrium systems can be characterised by
mean-field variables if field fluctuations are negligible
(fluctuations can however be introduced phenomenologically into the
evolution equations if required).  A mean-field approach leads
naturally to the concept of the order parameter, and the corresponding
order parameter equation.  The order parameter is either a physical
field or an abstract field which acquires a non-zero value in the an
ordered phase (such as a Bose-condensed or lasing state), and vanishes
in the normal state.  When considering a spatially inhomogeneous
system (with trapping, or inhomogeneous pumping), the order parameter
may vary in space. When considering a non-equilibrium system, or the
dynamics of a system as it approaches its equilibrium state, the order
parameter may also vary in time. In such cases, the order parameter
equation describes the space and time dependence of the order
parameter, accounting for the generic features of the system's
dynamics.

One important classification of order parameter equations
distinguishes whether they describe relaxation towards an equilibrium
configuration, or phase evolution in a conservative system, or a
mixture of the two\cite{PismenBook:1999}.  For a dissipative system,
the dynamics may be described by using an energy functional
$\mathcal{F}[\psi]$, written in terms of the order parameter $\psi$
and its spatial derivatives.  The dissipative system dynamics causes
this energy functional to decay as a function of time, reaching a
minimal value at equilibrium, i.e. $\partial_t \psi = -
\Gamma \partial_\psi \mathcal{F}$.  The dynamical critical behaviour
of such systems has been extensively reviewed by Hohenberg and
Halperin \cite{Hohenberg:1977}. Such an approach is appropriate for
many solid-state systems, including in particular non-equilibrium
superconductivity~\cite{kopnin:2001}.  In contrast, for isolated
systems such as ultracold atomic gases, the order parameter obeys
conservative dynamics, in which the energy functional does not change
with time, and the order parameter instead follows Hamiltonian
dynamics.  We will discuss the behaviour of systems that lie between
these two extremes of purely dissipative and purely conservative
dynamics, a scenario that includes the non-equilibrium polariton
condensate.

The structure of the energy functional, and thus of the resulting
order parameter equation, is determined by the symmetries of the order
parameter space.  Taking into account also the fact that near a phase
transition, the characteristic lengthscale of fluctuations diverges,
it becomes possible to restrict the form of the energy functional by
keeping the lowest order derivative terms that possess the required
symmetries.  This makes it possible to divide systems into
universality classes, depending only on the symmetries and the nature
of the dynamics~\cite{Hohenberg:1977,kadanoff:2000}.  Identifying
which classes various system belong to allows one to draw similarities
between systems that are very different in nature and to predict the
behaviour of the new systems that fall into previously known
universality class.  For instance, symmetry under changing the phase
of of the order parameter restricts the energy functionals to
dependence on $|\psi|$ only, and considering the lowest order form
that allows for a symmetry breaking from disordered (zero) to ordered
(nonzero) state gives a potential as a quartic polynomial in $|\psi|$.
Including the lowest compatible order of spatial derivative terms then
gives the energy functional whose dissipative dynamics correspond to
the Ginzburg-Landau equation:
 \begin{equation}
 {\cal{F}}=\int dV
 \nabla\psi \cdot \nabla \psi^* + (\mu - U_0|\psi|^2)^2, \label{berloff:L0}
\end{equation}
where $\mu$ is the control parameter that forces the system to move
from the normal state $|\psi|=0$ to the ordered state with
$|\psi|^2=\mu/U_0$.  Eq.~(\ref{berloff:L0}) is expected to be relevant
to a physical system in the vicinity of the phase transition, where the
smallness of the modulus and the derivatives of the order parameter
allow to keep only the leading order terms in the expansion. Higher
derivatives and other higher order terms can be kept to allow for more
complex forms of order and associated phase transitions.

Understanding the universality class of a given system leads to
understanding of fundamentals of the behaviour of that system.  The
studies of vortices and vortex dynamics in superfluid helium
\cite{Hall:1956eh} led to prediction and experimental realisation of
vortices first in nonlinear optics \cite{Arecchi:1991jw}, then in
atomic Bose-Einstein condensates \cite{Dodd:1997cj} and finally in
nonequilibrium solid-state condensates \cite{Liew:2007kn,
  Lagoudakis:2008ia} all due the hydrodynamic interpretation of the
order parameter equations.  Spiral waves in biological and chemical
systems suggested the existence of such meandering waves in class B
lasers. Solitary waves in atomic systems all have their analogs in
nonlinear optics. Finally, much of the experiments in solid-state
condensates are now motivated by finding localised excitations similar
to other system that share the same universality class
\cite{Amo:2011bfa,Dzyapko:2010tf}.  Pattern formation in systems that
belong to the same universality class share similar
properties. Patterns appear in open nonlinear systems when an
amplitude distribution of the order parameter becomes unstable above a
certain threshold. Linear instability gives rise to a so-called pure
state that, if nonlinearities are weak, can dominate the dynamics.
Strong nonlinearities may mix the eigenvalues leading to various
stationary or chaotic combinations of pure states with different
combinations occupying either all space or different space regions.

In this chapter we shall follow the evolution from equilibrium
condensates to non-equilibrium condensates to lasers analysing their
universality, emphasising similarities and differences. We will
discuss in some detail the origin of the most general order parameter
equation for the laser system, and comment on the relation of this
order parameter equation to that for cold atoms and for
non-equilibrium polariton condensation.  We will then demonstrate how
the various terms that may exist in the order parameter equation
affect the patterns which arise, focusing on three cases: the
case with homogeneous pumping and no trapping,  the case with
inhomogeneous pumping and no trapping, and the case with an harmonic
trap.

\subsection{Review of physical systems}

Laser dynamics is described by coupling Maxwell equations with
Shr\"odinger equations for $N$ atoms confined in the cavity.
Expanding the electric field in cavity modes and keeping only the
leading order mode leads to the equations that couple the amplitude of
this mode with the collective variables that describe the polarisation
and population of the gain medium.  Such coupled equations are known
as Maxwell-Bloch (MB) equations.  Lasers are then classified depending
on the relative order of the loss rates for the electric field,
compared to the decay rates of the gain medium polarisation and
population. The MB equations have two homogeneous stationary
solutions: nonlasing (zero order parameter) and lasing (nonzero order
parameter) solutions. The instabilities of these solutions, and
therefore, pattern formation, are described by universal order
parameter equations: the complex Swift-Hohenberg (cSH) equation for
lasers with a fast population inversion and the cSH equation coupled
to a mean flow if the population inversion is slowly varying. The
universal equation describing the bifurcation of lasing solution takes
the form of a cSH equation coupled to a Kuramoto-Sivashinsky equation
\cite{Kuramoto:1976,Sivashinsky:1977}.

Semiconductor microcavities confine photon modes, which may then
interact with electronic excitations in the semiconductor.  If the
cavity is resonant with the energy to create an exciton (a bound
electron hole pair), and if the exciton-photon coupling is strong
enough, then new normal modes (new quasiparticles) arise as hybrids of
excitons and photons: polaritons.  For low enough densities, these may
be considered as bosonic quasiparticles, and so can form a condensed
(coherent) state above a critical density.  These are intrinsically
non equilibrium systems with the steady states set by balance between
pumping and losses due to the short lifetime of polaritons. Depending
on whether the emission from the microcavity follows the bare photon
or the lower polariton dispersion the system shows either regular
lasing or polariton condensation and in this sense crosses over
continuously from weak coupling at higher temperatures and pumping
strengths to strong coupling at lower temperatures and lower pumping
intensities. Losses in the microcavity systems can be decreased by
improving the quality of dielectric Bragg mirrors. The smaller the
pumping and losses become the closer polariton condensates come to
resemble equilibrium Bose-Einstein condensates (BECs). It seems
therefore that the unified approach should be possible to describe the
transition from normal lasers to the equilibrium BECs via polariton
condensates. There are some other differences between atomic or
polariton condensates and normal lasers. The operation of a photon
laser is based on three ingredients: a resonator for the
electromagnetic field, an gain medium and a excitation mechanism for
the gain medium.  When excited, the gain medium will undergo
stimulated emission of radiation that amplifies the electromagnetic
field in the cavity. In contrast, for polariton condensates there is
instead stimulated scattering within the set of polariton modes, and
condensation can take place without any inversion of the gain medium,
and thus potentially having a lower threshold \cite{Imamoglu:1996vwa}.
A microscopic theory would be required to fully describe how all of
these aspects cross over from polariton condensation to lasing,
however given the universality of order parameter equations, one may
hope to write a single order parameter equation which captures these
different regimes by varying appropriate parameters.

\section{Derivation of order parameter equations}

In this section we show how various order parameter equations arise in
descriptions of lasers, and how these relate to the order parameter
equations relevant to ultracold atoms and polariton condensates.  We
start with a mean field (semiclassical) model of a laser, the MB
system of equations.  In section~\ref{sec:fast-reserv-deph} we show
how the assumption of small relaxation times for atomic polarisation
in comparison with the cavity relaxation time reduces these equations
to the complex Ginzburg-Landau (cGL) equations \cite{Aranson:2002kba}
or the coupled cGL equation and the gain medium population dynamics
\cite{Taranenko:1997bt}. These models have been extensively used to
model non-equilibrium condensates
\cite{Dzyapko:2010tf,Wouters:2010gwb,Keeling:2008hj,Wouters:2007dia}. We
discuss how the mode selection, in which a particular transverse mode
grows fastest, is lost in the derivation of these models. Then, in
Sec.~\ref{sec:multi-scale-analysis} we instead follow the derivation
used in \cite{Moloney:1995usa} based on the multi-scale expansion
technique to derive the cSH equations for class A and class C lasers
\cite{Cross:1993ela}.  In section~\ref{sec:modell-excit-polar} we then
discuss how nonlinear interactions appear in these equations, and
discuss the interpretation of these equations as order parameter
equations for polariton condensates. If the reservoir dynamics is slow
in comparison with time evolution of the order parameter these
equation should be replaced by a coupled system explicitly modelling
the reservoir dynamics.  In the limit of the long life-time of the
particles the system becomes the Gross-Pitaevskii (GP) equation aka
the nonlinear Schr\"odinger (NLS) equation that describes atomic BECs.

\subsection{Maxwell-Bloch equations for a laser}

We start with the MB equations for a wide-aperture laser with an
intracavity saturable absorber with multiple transverse modes in the
single longitudinal mode approximation~\cite{Lugiato:1988bta}
\begin{eqnarray} 
&&\frac{\partial E}{\partial t} - i  \nabla^2 E = P_g - P_a-(1 + i \Delta_e)E,\label{berloff:mb0a}\\
&&\tau_{\perp g}\frac{\partial P_g}{\partial t}+ (1+i\Delta_g)P_g= EG,\label{berloff:mb0b}\\
&&\tau_{\perp a} \frac{\partial P_a}{\partial t}+ (1+i\Delta_a)P_a= EA,\label{berloff:mb0c}\\
&&\tau_g \frac{\partial G}{\partial t} = G_0 - G -\frac{1}{2}(E^*P_g+EP_g^*),\label{berloff:mb0d}\\
&&\tau_a \frac{\partial A}{\partial t}  = A_0 - A - \frac{D}{2}(E^*P_a+EP_a^*),
\label{berloff:mb0e}
\end{eqnarray}
where the complex field $E$ is the envelope of the electric field, the
real functions $G$ and $A$ are the population differences for gain and
absorption media, the complex functions $P_g$ and $P_a$ are the
envelopes of polarisation for gain and absorption media. $G_0$
and $A_0$ are the stationary values of the population difference in
the absence of the laser field; they are proportional to the external
gain and losses in the system.The parameter $D=\tau_{\perp a} \tau_a
\mu_a^2/(\tau_{\perp g} \tau_g \mu_g^2)$ is the relative saturability
of gain and loss media and $\mu_a$ and $\mu_g$ stand for the atomic
dipole momenta. The parameters $\tau_{\perp a,g}$ and $\tau_{a,g}$ are
the relaxation times for atomic polarisations and population
differences scaled by the cavity relaxation time, the time $t$ is also
scaled by the cavity relaxation time. The parameters
$\Delta_a-\Delta_e=(\omega_a - \omega_c)\tau_{\perp a}$ and
$\Delta_g-\Delta_e=(\omega_g - \omega_c)\tau_{\perp g}$ are detunings
between the spectral line centre and the frequency of empty cavity
mode $\omega_c$.  Without loss of generality we work in rotating frame
such that the the fast time is eliminated via introduction of
$\Delta_e$. The spatial coordinates are rescaled by the width of the
effective Fresnel zone.

\subsection{Fast reservoir dephasing limit}
\label{sec:fast-reserv-deph}

Following \cite{Staliunas:1993fk,Fedorov:2000uja} we assume that $\tau_{\perp a}, \tau_{\perp g}$ are small and consider the first-order approximations to Eqs. (\ref{berloff:mb0a}-\ref{berloff:mb0e}). Keeping up to the linear terms in these small quantities  we get
\begin{equation}
P_g=\frac{G E}{1 + i \Delta_g}- \tau_{\perp g} \frac{(G E)_t }{(1 + i \Delta_g)^{2}},
\label{P}
\end{equation}
and similar for $P_a$. The equation on $E$ after we substitute these expressions for $P_g$ and $P_a$ becomes
\begin{equation}
(1+i \eta)\frac{\partial e}{\partial t} - i (\nabla^2-\Delta_e) e= [(1 - i \Delta_g) g - (1 - i\Delta_a) a-1]e,
\label{berloff:e}
\end{equation}
where 
\begin{equation}
\eta=-2 \tau_{\perp g} g \Delta_g/(1 + \Delta_g^2) + 2 \tau_{\perp a} a \Delta_a/(1 + \Delta_a^2) ,
\label{berloff:diffusion}
\end{equation}
and where we rescaled fields as $e=E/(1 + \Delta_g^2), g=G/(1 +
\Delta_g^2)$ and $a=A/(1 + \Delta_a^2)$. In writing Eq.
(\ref{berloff:e}) we kept leading order contributions in the imaginary
coefficient of time derivative (which is of order $O(\tau_{\perp g},
\tau_{\perp a})$). The real coefficient of the time derivative we kept
to $O(1)$ in $\tau_{\perp g}$ and $ \tau_{\perp a}$. The equations for
scaled gain and absorption media populations to  leading order take the forms:
\begin{eqnarray}
&&\tau_g\frac{\partial g}{\partial t} = g_0 - (1 + |e|^2) g,\label{berloff:g}\\
&&\tau_a\frac{\partial a}{\partial t} = a_0 - (1 + d |e|^2) a,\label{berloff:a}
\end{eqnarray}
where $a_0=A_0/(1 + \Delta_a^2), g_0=G_0/(1 + \Delta_g^2)$ and $d=D(1 + \Delta_g^2)/(1 + \Delta_a^2)$.

\subsubsection{Fast reservoir population relaxation}
The system of equations (\ref{berloff:e}, \ref{berloff:g},
\ref{berloff:a}) can be simplified under more stringent restrictions
on parameters.  In the limit of fast population relaxation times
$\tau_{g},\tau_{a} \ll 1$ (class A and C lasers) Eqs.( \ref{berloff:g},
\ref{berloff:a}) give
\begin{equation}
g=\frac{g_0}{1+|e|^2}, \qquad a=\frac{a_0}{1+d|e|^2},
\end{equation}
and Eq. (\ref{berloff:e})  becomes
\begin{equation}
(1 + i \eta(e))\frac{\partial e}{\partial t} - i \nabla^2 e=\biggl [\frac{(1 - i \Delta_g) g_0}{1+|e|^2} - \frac{(1 - i\Delta_a) a_0}{1+d|e|^2}-1\biggr]e,
\label{berloff:e2}
\end{equation}
where the  coefficient  $\eta$ is given by
\begin{equation}
\eta(e)=-2\biggl[\frac{ \tau_{\perp g} g \Delta_g}{(1 + \Delta_g^2) (1 + |e|^2)}-  \frac{\tau_{\perp a} a \Delta_a}{(1 + \Delta_a^2) (1 + d |e|^2)}\biggl].
\label{berloff:diffusion2}
\end{equation}

Close to the emission threshold $|e|^2 \ll 0$, which allows a cubic approximation for the nonlinear terms we get the complex Ginsburg-Landau equation (cGL) \cite{Aranson:2002kba}
\begin{equation}
(i - \eta(0))\frac{\partial e}{\partial t} = - \nabla^2 e+V e + U|e|^2e +i[\alpha-\beta |e|^2]e,
\label{berloff:e3}
\end{equation}
where we let $\alpha=g_0-a_0-1$, $\beta=g_0-a_0$, $U=d a_0\Delta_a-g_0\Delta_g$ , $V= g_0\Delta_g-a_0\Delta_a$.

The cGL equation is not a very accurate model of a laser since it does
not take into account the selection of transverse modes. The lasers
emit particular transverse modes that depend on the length of the
resonator. By making the assumption that $\tau_{\perp g},\tau_{\perp
  a}\rightarrow 0$, we assumed that the gain line is infinitely
broad. In order to take into account the tunability of lasers that
allows spatial-frequency selection a more careful derivation of the
order parameter equation is required, which does not take this limit
of fast polarisation relaxation.
\subsection{Multi-scale analysis of the Maxwell-Bloch equations}
\label{sec:multi-scale-analysis}
In this section we derive the complex Swift-Hohenberg equation capable
of selecting particular transverse modes from
the MB Eqs. (\ref{berloff:mb0a}--\ref{berloff:mb0e}). A similar
derivation has been done for the MB equations taking into account gain
only and assuming that $\Delta_g$ is small \cite{Lega:1994kaa}. Here we
shall only assume that $\nabla^2 - \Delta_e$ is small and use it as a
small parameter, $\epsilon(\nabla^2 - \Delta_e)$.  We apply the
technique of multi-scale expansion to $E,P_{g,a},G$ and $A$ looking
for solutions in the form of a power series expansion in $\epsilon$,
and introducing two slow time scales $T_1=\epsilon t$, $T_2=
\epsilon^2 t$, so that $\partial_t=\epsilon \partial_{T_1} +
\epsilon^2 \partial_{T_2}$. Next we solve equations at equal powers of
$\epsilon$. At the leading order we get non-lasing solution $(E,
P_g,P_a,G,A)=(0,0,0,G_0,A_0)$. At $O(\epsilon)$,
$(E_1,P_{g1},P_{a1},G_1,A_1)=(\psi, G_0 \psi/(1 + i \Delta_g),
A_0\psi/(1 + i\Delta_a), 0, 0)$, where $\psi$ is a yet undetermined
complex field and $G_0$ and $A_0$ are linked via $1=G_0/(1 +i\Delta_g)
- A_0/(1 + i\Delta_a)$. This condition specifies $G_0$ and $A_0$ at
the threshold for laser emission as $G_{\text{crit}}=\Delta_a(1 +
\Delta^2_g)/(\Delta_a-\Delta_g)$ and $A_{\text{crit}}=\Delta_g(1 +
\Delta^2_a)/(\Delta_a-\Delta_g)$. We make near-threshold assumption
$G_0=G_{\text{crit}}+ \epsilon^2 l_g$ and $A_0=A_{\text{crit}}+ \epsilon^2
l_a$. At $O(\epsilon^2)$ we get
\begin{eqnarray}
&&\frac{\partial \psi}{\partial T_1} = i (\nabla^2 - \Delta_e)\psi + P_{g2}-P_{a2} - E_2, \\
&&\tau_{\perp g} \frac{\partial P_{g1}}{\partial T_1} + (1 + i \Delta_g) P_{g2} = E_2 G_0,\\
&&\tau_{\perp a} \frac{\partial P_{a1}}{\partial T_1} + (1 + i \Delta_a) P_{a2} = E_2 A_0,\\
&&0=-G_2 - \frac{1}{2}(\psi P_{g 1}^* + \psi^*P_{g 1}),\\
&&0=-A_2 - \frac{D}{2}(\psi P_{a 1}^* + \psi^*P_{a 1}).
\end{eqnarray}
From these equations we get the compatibility condition
\begin{equation}
(1 + \widetilde{G_0}\widetilde{\tau}_{\perp g} - \widetilde{A_0}\widetilde{\tau}_{\perp a}) \frac{\partial \psi}{\partial T_1} = i (\nabla^2 - \Delta_e) \psi,\label{berloff:psi1}
\end{equation}
and expressions for $P_{g2},P_{a2},G_2$ and $A_2$
\begin{eqnarray}
&&P_{g2}=-\widetilde{\tau}_{\perp g} \widetilde{G_0}\frac{\partial \psi}{\partial T_1}, \quad P_{a2}=-\widetilde{\tau}_{\perp a} \widetilde{A_0}\frac{\partial \psi}{\partial T_1},\label{berloff:p2}\\
&&G_2=-\frac{G_0 |\psi|^2}{1 + \Delta_g^2}, \quad A_2=-\frac{A_0 D |\psi|^2}{1 + \Delta_a^2}, \label{berloff:ga}
\end{eqnarray}
where we let $E_2=0$ and  denoted $\widetilde{\tau}_{\perp g,\perp a}=\tau_{\perp g,\perp a}/(1 + i\Delta_{g,a})$, $\widetilde{G_0}=G_0/(1 + i\Delta_{g})$ and $\widetilde{A_0}=A_0/(1 + i\Delta_{a})$.
At $O(\epsilon^3)$ we get 
\begin{eqnarray}
&&\frac{\partial \psi}{\partial T_2}  =  P_{g3}-P_{a3} - E_3, \\
&&\tau_{\perp g} \biggl(\frac{\partial P_{g1}}{\partial T_2}+\frac{\partial P_{g2}}{\partial T_1}\biggr) + (1 + i \Delta_g) P_{g3} = E_3 G_0 + \psi (G_2+l_g),\\
&&\tau_{\perp g} \biggl(\frac{\partial P_{a1}}{\partial T_2}+\frac{\partial P_{a2}}{\partial T_1}\biggr) + (1 + i \Delta_a) P_{a3} = E_3 A_0 + \psi (A_2+l_a),\\
&&\tau_g\frac{\partial G_{2}}{\partial T_1}=-G_3 - \frac{1}{2}(\psi P_{g 2}^* + \psi^*P_{g 2}),\\
&&\tau_a\frac{\partial A_{2}}{\partial T_1}=-A_3 - \frac{D}{2}(\psi P_{a 2}^* + \psi^*P_{a 2}).
\end{eqnarray}
The compatibility condition at this order after we substitute (\ref{berloff:p2})--(\ref{berloff:ga}) gives
\begin{eqnarray}
&&(1 + \widetilde{G_0}\widetilde{\tau}_{\perp g} - \widetilde{A_0}\widetilde{\tau}_{\perp a}) \frac{\partial \psi}{\partial T_2}+\widetilde{\tau}_{\perp g}\frac{\partial P_{g2}}{\partial T_1} -\widetilde{\tau}_{\perp a}\frac{\partial P_{a2}}{\partial T_1}\nonumber\\ 
&&=\biggl(\frac{l_g}{1 + i \Delta_g}-\frac{l_a}{1 + i \Delta_a}\biggr)\psi -\biggl(\frac{\widetilde{G_0}}{1 + \Delta_g^2}-\frac{\widetilde{A_0}D}{1 + \Delta_a^2}\biggr)|\psi|^2\psi.\label{berloff:psi2}
\end{eqnarray}
We use Eqs. (\ref{berloff:psi1}) and (\ref{berloff:p2}) in Eq. (\ref{berloff:psi2}), collect the derivatives as $\partial_t = \epsilon \partial_{T_1} + \epsilon^2 \partial_{T_2}$,  absorb $\epsilon$ into $\psi$  and $\nabla^2 - \Delta_e$ and  replace  $\epsilon^2 l_{g}$ ($\epsilon^2l_{a}$) with $G_0-G_{\text{crit}}$ ($A_0-A_{\text{crit}}$) as expected. The result is the cSH equation
\begin{eqnarray}
(1 + \widetilde{G_0}\widetilde{\tau}_{\perp g} - \widetilde{A_0}\widetilde{\tau}_{\perp a}) \frac{\partial \psi}{\partial t}&=&i(\nabla^2 - \Delta_e)\psi - \frac{(\widetilde{\tau}^2_{\perp g}\widetilde{G_0} - \widetilde{\tau}^2_{\perp a}\widetilde{A_0})}{(1 + \widetilde{G_0}\widetilde{\tau}_{\perp g} - \widetilde{A_0}\widetilde{\tau}_{\perp a})^2} (\nabla^2 - \Delta_e)^2 \psi \nonumber \\
&+&\gamma \psi- \biggl(\frac{\widetilde{G_0}}{1 + \Delta_g^2}-\frac{\widetilde{A_0}D}{1 + \Delta_a^2}\biggr)|\psi|^2\psi,
\label{berloff:csh0}
\end{eqnarray}
where $\gamma=(G_0-G_{\text{crit}})/(1 + i\Delta_g) - (A_0-A_{\text{crit}})/(1 + i\Delta_a)$.

We can simplify the coefficients by considering a limit $\Delta_{g,a} \ll \tau_{\perp g,a} \ll 1$, neglecting $O(\Delta_{g,a}^2)$ and $O(\tau_{\perp g,a}^2 \Delta_{g,a})$ terms and keeping only the higher order terms for real and imaginary parts of the coefficients. This leads to the following general form of the cSH equation
\begin{eqnarray}
(1 + i \eta) \frac{\partial \psi}{\partial t}&=&i(\nabla^2 - \Delta_e)\psi -  \delta(\nabla^2 - \Delta_e)^2 \psi \nonumber \\
&+&(\alpha - i V) \psi- (\beta+ i U)|\psi|^2\psi,
\label{berloff:cshfin}
\end{eqnarray}
with the energy relaxation $\eta=-2 G_0\Delta_g\tau_{\perp g}+ 2 A_0\Delta_a\tau_{\perp a}$, the coefficient of superdiffusion $\delta=\tau_{\perp g}^2 G_0 - \tau_{\perp a}^2 A_0$, the  effective pumping $\alpha=G_0-A_0-1$, the effective repulsive potential $V=G_0\Delta_g-A_0\Delta_a$, the cubic damping $\beta=G_0-A_0D$ and interaction potential $U=A_0 D \Delta_a - G_0 \Delta_g$.

Apart from nonlinear optics and lasers the cSH equation provides a reduced description of a variety of other systems \cite{Cross:1993ela}, such as Rayleigh-Bernard convection \cite{Cox:2004baa},  Couette flow \cite{Manneville:2004vfa}, nematic liquid crystal \cite{Buka:2004tja}, magnetoconvection \cite{Cox:2004baa} and propagating flame front  \cite{Matkowsky:2003wba} among others. 

Similar to other universal equations the cSH equation can be derived
phenomenologically from general symmetry considerations. Assuming that
the system is characterised by an instability at $k_c\ne 0$, the
dominating growth rate (Lyapunov exponent) can be approximated close
to $k_c$ by a parabola that takes positive values in the neighbourhood
of $k_c$. To the lowest degree of approximation this can be modelled
by
\begin{equation}
\lambda= \alpha-\delta(k^2  - k_c^2)^2 + i  ( k^2 - k_c^2),
\end{equation}
where $\alpha$ is a control parameter that takes $Re(\lambda)$ into the positive range of values. A linear model that has the corresponding dispersion has to be complemented with 
a  nonlinear term in order to prevent the infinite growth of unstable modes. The simplest form of such nonlinearity that preserves the  invariance of the field phase is the cubic nonlinearity $|\psi|^2\psi$.  So the minimum equation that describes a class of phenomena in nonlinear optics in the lowest order approximation coincides with the cSH equation (\ref{berloff:cshfin}).

\subsubsection{Slow population  evolution}
For wide aperture CO${}_2$ and semiconductor lasers the cSH equation introduced in the previous section is not a good model. The population  dynamics is slow which corresponds to the case of the stiff MB equations that occurs when the parameter $b=\tau_{\perp g,a}/\tau_{g,a}$, that measures the ratio of the polarisation dephasing to the population deenergisation rate, becomes small. The order parameter equation in this case is not a single equation and the analysis of the previous section should be revised taking into account smallness of $b$ \cite{Lega:1994kaa}.
Instead of going through the multi-scale analysis we note that we can consider gain selection separate from population evolution and therefore rewrite Eqs. (\ref{berloff:e},\ref{berloff:g},\ref{berloff:a}) to include the gain selection  mechanism
\begin{eqnarray}
(1 + i \eta) \frac{\partial \psi}{\partial t}&=&i(\nabla^2 - \Delta_e)\psi -  \delta(\nabla^2 - \Delta_e)^2 \psi -\psi \nonumber \\
&+&[(1 - i \Delta_g) G - (1 - i \Delta_a)A]\psi,
\label{berloff:esch}\\
&&\tau_g\frac{\partial G}{\partial t} = G_0 - (1 + |\psi|^2) G,\label{berloff:gsch}\\
&&\tau_a\frac{\partial A}{\partial t} = A_0 - (1 + D |\psi|^2) A.\label{berloff:asch}
\end{eqnarray}
One may note that in the limit that $\tau_g, \tau_a$ are small, this
equation reduces to Eq.~(\ref{berloff:cshfin}). 

\subsection{Modelling exciton-polariton condensates}
\label{sec:modell-excit-polar}

The cSH order parameter equation derived above from the MB equations
of a laser can also describe the polariton condensate.  In this
section we discuss how such an equation can arise for the polariton
system, and the meaning the various terms would acquire in this
context.  We also make contact with the limiting cases which
correspond to ultracold atomic gases.  For the polariton condensate we
interpret $\psi$ in Eqs. (\ref{berloff:esch}--\ref{berloff:asch}) as a
scalar mean field of a polariton matter-wave field operator
$\widehat{\Psi}({\bf r},t)$.  We begin by considering the basic energy
functional for a polariton condensate.  In addition to the kinetic
energy, and any external trapping potential, one must also take into
account repulsive interactions of polaritons.  These interactions
predominately come from the short ranged electron--electron exchange
interactions (when two excitons swap their electrons).  This
interaction gives rise to a cubic nonlinear term $-iU_0|\psi|^2\psi$
just as in the right-hand side of Eq.  (\ref{berloff:cshfin}).  Rather
than coupling the order parameter equation to the dynamics of the gain
medium, one should instead consider coupling of the order parameter
equation to the equation describing the density of noncondensed
polaritons (reservoir excitons), $G$
\cite{Kneer:1998fma,Wouters:2007dia}, that may also contain a
diffusion term.

In the limit of vanishing gain and losses, the order parameter
equation becomes the NLS equation also
used to model a Bose-Einstein
condensation of ultracold atoms:
\begin{equation}
\frac{\partial \psi}{\partial t} =i \nabla^2\psi
+ i V(r) \psi
 -iU_0|\psi|^2\psi.
\label{berloff:nls}
\end{equation}
For an ultracold atomic gas this equation can also be derived
microscopically from the Heisenberg representation of the many-body
Hamiltonian using the language of second quantisation.
For the case of an ultracold atomic gas, one may also include the
dissipation that arises from collisions of condensate atoms with non
condensed thermal cloud in this equation.  This process leads to
energy relaxation and atom transfer between the condensate and the
thermal cloud.  This can be modelled by writing the quantum Boltzmann
equation, i.e. kinetic equation, describing the dynamics of the
populations of states \cite{Penckwitt:2002kda,Gardiner:1997jca}.  The
net rate of atom transfer $\eta$ as the result of such collision can
be represented by replacing the time derivative in
Eqs. (\ref{berloff:nls}) as $\partial_t \rightarrow
(1+i\eta)\partial_t$.  This parameter $\eta$ depends on the
temperature and the density of the noncondensed cloud.  Similar
mechanism of energy relaxation exists in polariton condensates and
have been phenomenologically introduced into various models of
polariton condensates \cite{Wouters:2010gwb,Wouters:2010ee}. Note that
such energy relaxation follows directly from the MB equations as
indicated by Eqs. (\ref{berloff:e}, \ref{berloff:cshfin}). The interactions with noncondensed
cloud may enhance this coefficient.

In addition to the terms mentioned so far, the polariton system
differs from ultracold atoms, but is similar to the laser, in having
also terms describing gain and loss, i.e. pumping and decay.
Including these terms, and allowing them to potentially depend on
wavevector, gives a modified cSH model that includes all possible
previously discussed limits of lasers, nonequilibrium polariton
condensates and equilibrium atomic BECs:
\begin{eqnarray}
(1 + i \eta) \frac{\partial \psi}{\partial t}&=&i(\nabla^2 - \Delta_e)\psi -  \delta(\nabla^2 - \Delta_e)^2 \psi -\psi \nonumber \\
&+&[(1 - i \Delta_g) G - (1 - i \Delta_a)]\psi-i U_0|\psi|^2\psi,
\label{berloff:ee}\\
&&\tau_g\frac{\partial G}{\partial t} = G_0 - (1 + |\psi|^2) G,\label{berloff:gg}\\
&&\tau_a\frac{\partial A}{\partial t} = A_0 - (1 + D |\psi|^2) A,\label{berloff:aa}
\end{eqnarray}
Note that $V=G\Delta_g-A\Delta_a$ gives rise to a reservoir potential
which causes the blue-shift in the condensate \cite{Tosi:2012ika}.

Some limiting cases of Eqs. (\ref{berloff:ee}--\ref{berloff:aa}) have
been previously considered.  Assuming $\delta\rightarrow 0$ and fast
relaxation of reservoirs ($\tau_a, \tau_g \to 0$) leads to the cGL
equation introduced for polariton condensates in
\cite{Keeling:2008hj}.  In the limit $\delta\rightarrow 0$ and
assuming the slow relaxation of the noncondensed reservoir gives rise
to the model of atom laser \cite{Kneer:1998fma} that has proved
effective for polariton condensates
\cite{Wouters:2007dia,Krizhanovskii:2009cea,Roumpos:2010kta,Wouters:2010gwb}.
Finally, in the limit of vanishing losses and gain all systems
approach the conservative NLS equation.

\section{Pattern formation and stability}

Having discussed the physical origin of the order parameter equations
of polariton condensates, lasers and atomic condensates, this section
now discusses the consequences of the form of the order parameter
equations for pattern forming and stability analysis.  We discuss
three cases: the entirely homogeneous case, the case in which the
pumping (gain) is localised, and the case in which there is
inhomogeneity of the condensate mode energy (i.e. trapping) as well as
pumping.  The homogeneous case is most relevant to wide aperture
lasers with electrical pumping.  For polariton condensates and photon
condensates with external pumping, the second and third cases are more
relevant.  As one goes toward equilibrium systems (such as atomic
condensates), the role of trapping potentials to confine the
condensate becomes more important, and so the third case is most
relevant in this limit.  

All three cases can be written as the short population relaxation time
limit of Eq.~(\ref{berloff:ee}):
\begin{eqnarray}
  (1+i\eta)\frac{\partial\psi}{\partial t} &=&(\alpha(r)-\beta|\psi|^2)\psi 
  + i ( \nabla^2 -V(r)-U_0|\psi|^2)\psi \nonumber \\
  && + 2\delta\Delta_e \nabla^2\psi - \delta\nabla^4\psi, \label{berloff:csh2}
\end{eqnarray}
but we will rescale the equation in different ways for the different cases.

\subsection{Behaviour of  homogeneous order parameter equation}

We begin by reviewing the simplest case, of linear stability analysis
about the uniform solution $\rho=\psi_0^2=\alpha/(\beta + U_0 \eta)$ in
Eq. (\ref{berloff:csh2}).  This uniform solution should be a stable solution as
long as $\delta \Delta_e < 0$.  One may then consider perturbations of
the Bogoliubov--de Gennes form:
\begin{equation}
\psi=
\left(
  \psi_0 +
  u e^{i({\bf k\cdot x} - \omega t)} 
  + 
  v^* e^{-i({\bf k\cdot x} - \omega^\ast t)}
\right)
e^{-i \mu t}
\end{equation}
where the chemical potential is $\mu = U_0 \rho$.  For this ans\"atz
to solve Eq.~(\ref{berloff:csh2}) (at linear order order in $u,v$)
requires that:
\begin{equation}
\label{berloff:spectrum}
\rho^2 (U_0^2 + \beta^2) =|S|^2 + \omega(1 - i \eta) S^* - \omega(1+i \eta) S - \omega^2 (1 + \eta^2),
\end{equation}
where $S=(k^2 + \rho U_0) + i (2 \delta \Delta_e k^2 + \delta k^4 +
\rho \beta)$ (making use of the steady state values of $\mu$ and
$\rho$). For an equilibrium condensate $\alpha=\beta=\delta=\eta=0$
one can recover the expected Bogoliubov spectrum from
Eq.~(\ref{berloff:spectrum}):
\begin{equation}
\omega_B(k)=\sqrt{k^2(2 \rho U_0 + k^2)}.
\end{equation}
Alternatively, in the cGL regime ($\delta=0$) with $\eta=0$ one recovers the
dissipative spectrum obtained previously \cite{Wouters:2008vq}:
\begin{equation}
  \omega_{cGL}(k)=-i \alpha \pm \sqrt{\omega_B(k)^2 -\alpha^2},
\end{equation}
which is imaginary for small $k$, and then above a critical $k$ (set by
$\omega_B(k) = \alpha$), the imaginary part becomes a constant $-\alpha$
and a real part appears.  Introducing the remaining terms gives
\begin{equation}
  \omega_{cSH}(k)=\frac{1}{1+\eta^2} \left[
    - i (\alpha + \chi_k)
    \pm \sqrt{- \alpha^2 + \epsilon_k (2 \rho(U_0 - \eta \beta) + \epsilon_k) }
  \right]
\end{equation}
where $\chi_k = k^2[ \eta + \delta (2\Delta_e + k^2)]$ and $\epsilon_k =
k^2[1 - \eta \delta (2\Delta_e + k^2)]$.  Note that for $k=0$, one
always has a mode at zero frequency, as expected given the phase
symmetry breaking present in the ordered phase.  As long as $\eta>0,
\delta\Delta_e>0$, the imaginary part grows for large $k$, since such a
case describes pumping that suppresses high energy (momentum) modes.
If $\eta +2 \delta  \Delta_e > 0 $, the modes are always decaying, but
if $\Delta_e < - \eta/2 \delta$, it becomes possible for the CSH term to
make the uniform part unstable --- the exact critical $\Delta_e$ depends
in a non-trivial way on the remaining parameters.  Other instabilities
may also arise due to the content of the square root term.

\subsection{Inhomogeneous pumping}
We next consider the effect of inhomogeneous pumping, comparing
the behaviour of  cSH equation and cGL equations
when used to model polariton condensates. As the first example we
consider a small pumping spot. This geometry has been considered
extensively in experiments
\cite{Wertz:2009jka,Tosi:2012ika,Christmann:2012wc,Roumpos:2010kta}
and theory \cite{Wouters:2010gwb}.  Our starting point is to consider
Eq. (\ref{berloff:csh2}) that we rewrite as
\begin{eqnarray}
  (1+i\eta(P))\frac{\partial\psi}{\partial t} &=&\biggl(P({\bf r})-\gamma_c-\lambda P({\bf r})|\psi|^2\biggr)\psi + i ( \nabla^2 -V(P)-|\psi|^2)\psi \nonumber\\
  &&+2\delta \Delta_e \nabla^2\psi - \delta \nabla^4\psi, \label{cSH3}.
\end{eqnarray}
We take
\begin{eqnarray} 
&&P({\bf r})= 4 \exp(-0.05 r^2), \qquad \gamma_c=0.3, \qquad \lambda=0.075, \nonumber\\
&&\eta(P)=0.025 P({\bf r}), \qquad V(P)=1.25 \exp(-0.45 r^2) P({\bf r}).
\label{data}
\end{eqnarray}
In writing the last two expressions we recalled that $\eta$, the
energy relaxation parameter representing the rate of transfer between
the noncondensed and condensed polaritons, depends on the density of
the noncondensed cloud. We also assumed spatially dependent energy
shifts coming from strong mutual repulsion \cite{Kasprzak:2006jy}, so
that the repulsive force coming from potential $V$ varies with density
of the condensate.

\begin{figure}[t]
\centering
\includegraphics*[width=.9\textwidth]{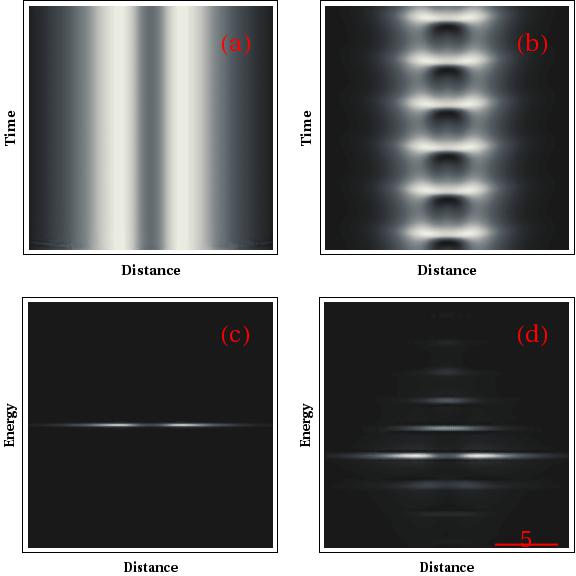}
\caption[]{Comparison between solutions of the cGL and cSH equations
  for a single pumping spot. Time evolution of the density measured
  across the pumping spot (a) for the cGL equation and (b) cSH
  equation. Energy of the solutions across the pumping spot (c) for
  the cGL equation and (d) cSH equation. The results of the numerical
  simulations of Eq. (\ref{cSH3}) with the parameters given in
  (\ref{data}).  }
\label{berloff:fig1}       
\end{figure}

We compare two cases: the cGL equation by letting $\delta=0$ in
Eq. (\ref{cSH3}) and the cSH equation with $\delta=0.1,
\Delta_e=-1$. In the case of the evolution according to the cGL
equation the system reaches the steady state, see
Fig.\ref{berloff:fig1}(a) which shows on the tomography image Fig
\ref{berloff:fig1}(c) as a single energy level. The evolution
according to the cSH equation leads to  periodic oscillations of
the density profile shown on Fig. \ref{berloff:fig1}(b). The
corresponding tomography image on Fig. \ref{berloff:fig1} shows
several discrete energy levels. Similar behaviour has been observed in
some experiments, eg. \cite{Roumpos:2010kta}.

\subsection{Inhomogeneous energy (trapping)}

We now turn to consider the behaviour in the presence of an harmonic
trap~\cite{Balili:2007gca,Klaers:2010bi}.  We will consider how the
presence of the dissipative terms in the general order parameter
equation affects the stability of known solutions of the Gross
Pitaevskii equation.  We will look both at linear stability analysis
(where one can gain insights from analytical results found by
considering the perturbative effect of dissipation), as well as full
numerical time evolution to find the new steady states.

As a starting point in the absence of dissipative terms, the
Gross-Pitaevskii equation:
\begin{equation}
  \label{berloff:gpe}
  \partial_t \psi 
  = i (\nabla^2 -r^2 - U_0|\psi|^2) \psi
\end{equation}
can be approximately solved by the stationary Thomas-Fermi profile
with $\partial_t \psi = -i\mu \psi$ and $|\psi|^2 =
\Theta(\mu-r^2)(\mu - r^2)/U_0$. This density profile results from
neglecting the kinetic energy.  This is valid as long as the cloud
size $r_{TF} = \sqrt{\mu}$ is large compared to the healing length
$1/\sqrt{U_0}$, i.e. for $\sqrt{U_0 \mu} \gg 1$. This stationary
Thomas-Fermi profile gives a simple prescription for how to find the
density profile in a given potential landscape.  However, as will be
discussed below, the stationary profile does not necessarily remain
stable in the presence of the additional terms in
Eq.~(\ref{berloff:csh2}).

\subsubsection{Stability analysis}

Starting from Eq.~(\ref{berloff:csh2}) with $\alpha, \beta, \delta, \eta =0$,
we consider in turn the effects introduced by adding these dissipative
terms.  We restrict to considering $\Delta_e < 0$ and neglect the 
superdiffusion term; after rescaling parameters, we
may write:
\begin{equation}
  \label{berloff:rescale-trap}
  2 {\partial_t \psi}
  -
  i \left( \nabla^2 - r^2 - |\psi|^2 \right) \psi
  =
  \left[
    \atil - \btil |\psi|^2 + 
    \dtil (  2  \Dtile - \nabla^2) \nabla^2 
    - 2 i \etil {\partial_t} 
  \right]
  \psi
\end{equation}
in which all dissipative terms are placed on the right hand side.  We
then proceed by considering normal modes around the stationary
solution in an approximation where the quantum pressure terms can be
neglected, this is done by writing the equations in terms of density
and phase and neglecting all quantum-pressure-like terms:
\begin{gather}
  \label{berloff:madellung}
  \partial_t \rho + \nabla\cdot(\rho \nabla \phi)
  =
  \left( \atil -\btil \rho + 2 \etil \partial_t \phi
    - 2 \dtil (\nabla \phi)^2 \right) \rho,
  \\
  2 \partial_t \phi +
  (\nabla \phi)^2 + r^2 + \rho =  \dtil
  (2\Dtile \nabla^2 - \nabla^4 )\phi.
\end{gather}
In the absence of the dissipative terms, this problem is the
two-dimensional analog of that studied by
Stringari~\cite{Stringari1996,Stringari1998}.  Linearising these
equations using $\rho \to \rho + h e^{-i\omega t}, \phi \to \phi +
\varphi e^{-i\omega t}$ yields normal modes with frequencies
$\omega_{ns} = \sqrt{2n^2 + 2(s+1)n + s}$ and density profiles given
by hypergeometric functions $h(r,\theta) \propto {}_2F_1(-n, n+s+1;
s+1, r^2) e^{i s \theta} r^s$; here $n$ is a radial quantum number,
and $s$ is an angular quantum number.  Including dissipative terms,
these normal mode frequencies acquire imaginary parts, describing
either growth or decay of such fluctuations.  Instability of the
stationary state occurs when at least one of these fluctuation modes
grows.

To account for dissipative terms perturbatively, it is enough to take
the normal mode functions found in the absence of dissipation and find
the first order frequency shift induced by the dissipative terms.  At
first order in the dissipative terms, there is no change to the
density profile; however a non-zero phase gradient $\nabla \phi$ does
appear at first order in the dissipative terms.  Vanishing of the
current at the edge of the cloud then requires $\mu = 3 \atil / (2
\btil + 3 \etil)$.

Formally one may write the linearised form of Eq.~(\ref{berloff:madellung})
in the form
\begin{math}
  -i\omega_{s,n} \chi_{s,n}(r,\theta) = 
  \left(\mathcal{L}^{(0)} + \mathcal{L}^{(1)}\right) \chi_{s,n}
\end{math}
in which $\mathcal{L}$ is a matrix of differential operators acting on
the fluctuation term $\chi = (h, \varphi)^T$.  By identifying the
dissipative terms as $\mathcal{L}^{(1)}$, standard first order
perturbation theory\footnote{ Some care must be taken since the
  operator $\mathcal{L}$ is not self adjoint and so the left and right
  eigenstates of $\mathcal{L}^{(0)}$ must  be found separately.  This
  is easiest if one replaces the variable $\varphi$ by $u_r
  = \partial_r \varphi, u_\theta = (1/r) \partial_\theta \varphi =
  (is/r) \varphi$, in this case the right eigenstates
  $(h,u_r,u_\theta)$ corresponds to the right eigenstate $(h,2 \rho
  u_r, 2 \rho u_\theta)$.}  then yields the first order correction:
$\omega^{(1)}_{ns} = i \langle \chi_{ns}^{(0)}, \mathcal{L}^{(1)}
\chi_{ns}^{(0)} \rangle / \langle \chi_{ns}^{(0)}, \chi_{ns}^{(0)}
\rangle$ where angle brackets indicate the appropriate inner product.
Following this procedure, one eventually finds
\begin{equation}
  \label{berloff:om-shift}
  \omega^{(1)}_{ns}
  = \frac{i}{2N} \int 2 \pi r dr
  \left[ (h_{ns}^{(0)})^2 \left(\atil - \etil \mu - (2 \btil + \etil) \mu \right)
    + \dtil h_{ns}^{(0)} \left(\Dtile-\frac{1}{2}\nabla^2\right)
    \nabla^2 h_{ns}^{(0)} \right]
\end{equation}
where the normalisation $N = \int 2 \pi r dr h_{ns}^2 $ and
integration is over the area of the Thomas-Fermi profile $r^2<\mu$.
The hypergeometric form of the zero order functions $h^{(0)}_{ns}$
allows Eq.~(\ref{berloff:om-shift})  to be evaluated analytically.
The terms proportional to $\dtil$ in fact vanishes, and the remaining
term can be written (making use of the above value of $\mu$) as:
\begin{equation}
  \label{berloff:om-shift-no-d}
  \omega^{(1)}_{ns}
  = \frac{i \atil}{4 \btil + 6 \etil} \left[
    (6 \btil + 3 \etil)
    \left( 
      \frac{s^2 + (\omega_{ns}^{(0)})^2}{s^2 + 2 (\omega_{ns}^{(0)})^2}
    \right)
    - 4 \btil - 3 \etil
  \right].
\end{equation}
Crucially, $(\omega_{ns}^{(0)})^2$ as given above grows only linearly
with $s$.  Thus, at large $s$ the ratio in parentheses tends to one,
and so
\begin{math}
  \omega^{(1)}_{ns}
  \to {i  \btil \atil}/(2 \btil + 3 \etil) > 0
\end{math}
This positive value means that there is an instability, even for
non-zero $\etil$.  Although neither $\etil$ nor $\dtil$ remove the
instability in this perturbative approach, this does not prevent these
terms from restoring stability via higher order corrections. This
needs to be checked by numerical simulations.
  
\subsubsection{Vortex lattices}
\begin{figure}[h]
\centering
\includegraphics*[width=.9\textwidth]{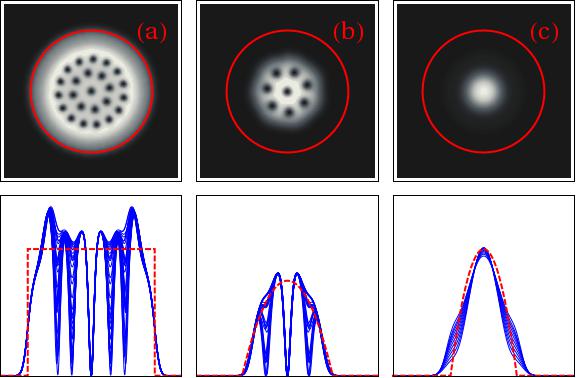}
\caption{Density plots of the polariton condensate in a harmonic trap
  according to the cGL equation with $\eta=0$ (a), $\eta=0.2$ (b) and
  the cSH equation with $\delta=\eta=0.2$ and $\Delta_e=-0.5$ (c). Top
  panels: luminosity is proportional to density. Red circles show the
  size of the pumping spot. Bottom panels: (solid lines) density of the cross
  section of the condensate at various times;  (dashed lines) analytic
  approximations, given for (a) and (b)  by Eq. (\ref{berloff:eq2}),
  and for (c) by the Thomas-Fermi profile $\Theta(\mu-r^2)(\mu-r^2)$
  with $\mu=3 \tilde\alpha_0/(2 \tilde \beta+ 3 \tilde \eta)$.  }
\label{berloff:fig2}       
\end{figure}

Having seen that the stationary profile possesses an instability, we
next consider the behaviour resulting from this instability.  In order
to reach a final configuration, it is necessary to restrict the
pumping to a finite range.  We thus take and $\atil({\bf r})=\atil_0
\theta(r_0-r)$, $\atil_0=8$, $\btil=0.6$, where the pumping is a flat
top of the radius $r_0=7$.  For $\etil = \dtil = 0$, this model has
been found to evolve to a rotating vortex
lattice\cite{Keeling:2008hj}. If one also includes the superdiffusion
present in the cSH model, one finds that (in contrast to the
linearised analysis) this may arrest the instability to vortex
formation, and thus lead instead to an oscillating vortex-free
state. Fig. \ref{berloff:fig2} compares the profiles that result from
the numerical simulation of Eq. (\ref{cSH3}) for the cases of the cGL
equations for $\etil=0$ (Fig. \ref{berloff:fig2}(a)), $\etil=0.2$
(Fig. \ref{berloff:fig2} (b)) and the cSH equation with
$\etil=\dtil=0.2$ and $\Dtile=-0.5$ (Fig. \ref{berloff:fig2} (c)).

Although the presence of $\etil$ does not remove the instability, it
does significantly effect the resulting rotating profile.  This can
both be seen in the numerical results shown in
Fig.~\ref{berloff:fig2}, and can also be understood by considering the
$\dtil=0$ limit of Eq.~(\ref{berloff:madellung}), written in a
rotating frame.  In a rotating frame, we consider solutions
to Eq.~(\ref{berloff:rescale-trap}) of the form:
\begin{equation}
  2 i \frac{\partial }{\partial t} \psi = \left(
    \mu - 2i {\bf \Omega}  \cdot  \vec{r} \times \nabla
  \right) \psi
\end{equation}
such that the time dependence has two parts: rotation with angular velocity ${\bf \Omega}=(0,0,\Omega)$, and phase
accumulation at rate $\mu$.  Inserting this ans\"atz into
Eq.~(\ref{berloff:rescale-trap}) and then making the Madellung
transform, with neglect of quantum pressure terms, one finds:
\begin{gather}
  \nabla \cdot \left[
    \rho (\nabla \phi - {\bf \Omega} \times \vec{r})
  \right]
  =
  \left[\atil - \btil \rho - \etil
    \left(
      \mu + 2 {\bf \Omega} \times \vec{r} \cdot \nabla \phi
    \right)
  \right] \rho
  \\
  -\mu +
  (\nabla \phi - {\bf\Omega} \times \vec{r})^2
  + (1-\Omega^2) r^2
  + \rho
  =0
\end{gather}
These equations can be satisfied by setting $\nabla \phi \simeq {\bf\Omega} \times \vec{r}$ which yields:
\begin{equation}
  \label{eq:1}
  0
  =
  \atil - \btil \rho - \etil
    \left(
      \mu + 2 \Omega^2 r^2
    \right),
  \qquad
  - \mu +  (1-\Omega^2) r^2 + \rho = 0
\end{equation}
These give two equations for $\rho$ which are both satisfied if:
\begin{equation}
  \label{berloff:eq2}
  \rho = \mu - (1-\Omega^2) r^2 = \frac{1}{\btil}\left[
    \atil - \etil(\mu + 2 \Omega^2 r^2) \right]
\end{equation}
hence $\Omega^2 = \btil/(\btil + 2 \etil), \mu=\atil/(\btil + \etil)$.
This indicates that while for $\etil=0$, the lattice rotates at
$\Omega=1$, cancelling out the trapping potential, for finite $\eta$,
the rotation velocity decreases, hence the density profile becomes
non-flat, as seen in Fig.~\ref{berloff:fig2}

In the above, $\nabla \phi \simeq {\bf \Omega} \times \vec{r}$ would require
the phase profile to mimic solid body rotation.  For a condensate,
this cannot be exactly satisfied, but can be approximately satisfied
(on a coarse grained scale) by having a density of vortices $\Omega$.
Since increasing $\etil$ causes $\Omega$ to decrease, a sufficiently
large value of $\etil$ can in effect kill any finite vortex lattice by
reducing the vortex density to values so that the number of vortices
falls below one.

\section{Conclusions}
We reviewed the connection between lasers, polariton condensates and
equilibrium Bose condensates from a common framework based on order
parameter equations. The cSH equations derived for lasers should be
applicable to polariton condensates in the limit of non-negligible
interactions and the stimulated scattering between polariton
modes. The pattern formation in the framework of the cSH equations
have been well-studied for lasers indicating a wealth of dynamics and
phenomena. Some of these phenomena may be achieved in polariton
condensates. At the same time the stronger nonlinearities and
different external potentials (engineered or due to disorder) may lead
to novel properties of the system exhibiting effects not seen in
normal lasers.

 \bibliographystyle{plain}
 \bibliography{csh}

\begin{thebibliography}{10}

\bibitem{Amo:2011bfa}
A~Amo, S~Pigeon, D~Sanvitto, V~G Sala, R~Hivet, I~Carusotto, F~Pisanello,
  G~Lemenager, R~Houdre, E~Giacobino, C~Ciuti, and A~Bramati.
\newblock {Polariton Superfluids Reveal Quantum Hydrodynamic Solitons}.
\newblock {\em Science}, 332(6034):1167--1170, 2011.

\bibitem{Aranson:2002kba}
Igor Aranson and Lorenz Kramer.
\newblock {The world of the complex Ginzburg-Landau equation}.
\newblock {\em Rev. Mod. Phys.}, 74(1):99--143, 2002.

\bibitem{Arecchi:1991jw}
F~Arecchi, G~Giacomelli, P~Ramazza, and S~Residori.
\newblock {Vortices and defect statistics in two-dimensional optical chaos}.
\newblock {\em Phys. Rev. Lett.}, 67(27):3749--3752, 1991.

\bibitem{Balili:2007gca}
R~Balili, V~Hartwell, D~Snoke, L~Pfeiffer, and K~West.
\newblock {Bose-Einstein Condensation of Microcavity Polaritons in a Trap}.
\newblock {\em Science}, 316(5827):1007--1010, 2007.

\bibitem{Buka:2004tja}
A~Buka, B~Dressel, and L~Kramer.
\newblock {Direct transition to electroconvection in a homeotropic nematic
  liquid crystal}.
\newblock {\em Chaos}, 14(3):793, 2004.

\bibitem{Christmann:2012wc}
G~Christmann, G~Tosi, N~G Berloff, P~Tsotsis, P~S Eldridge, Z~Hatzopoulos, P~G
  Savvidis, and J~J Baumberg.
\newblock Polariton ring condensates and sunflower ripples in an expanding
  quantum liquid.
\newblock arXiv:1201.2113, 2012.

\bibitem{Cox:2004baa}
S~M Cox, P~C Matthews, and S~L Pollicott.
\newblock {Swift-Hohenberg} model for magnetoconvection.
\newblock {\em Phys. Rev. E}, 69:066314, 2004.

\bibitem{Cross:1993ela}
M~Cross and P~Hohenberg.
\newblock {Pattern formation outside of equilibrium}.
\newblock {\em Rev. Mod. Phys.}, 65(3):851--1112, 1993.

\bibitem{Dodd:1997cj}
R~Dodd, K~Burnett, M~Edwards, and C~W Clark.
\newblock {Excitation spectroscopy of vortex states in dilute Bose-Einstein
  condensed gases}.
\newblock {\em Phys. Rev. A}, 56(1):587--590, 1997.

\bibitem{Dzyapko:2010tf}
O~Dzyapko, V~E Demidov, and S~O Demokritov.
\newblock {Ginzburg-Landau model of Bose-Einstein condensation of magnons}.
\newblock {\em Phys. Rev. B}, 81:024418, 2010.

\bibitem{Fedorov:2000uja}
S~V Fedorov, A~G Vladimirov, and G~V Khodova.
\newblock {Effect of frequency detunings and finite relaxation rates on laser
  localized structures}.
\newblock {\em Phys. Rev. E}, 61:5814, 2000.

\bibitem{Gardiner:1997jca}
C~Gardiner, P~Zoller, R~Ballagh, and M~Davis.
\newblock {Kinetics of Bose-Einstein Condensation in a Trap}.
\newblock {\em Phys. Rev. Lett.}, 79(10):1793--1796, 1997.

\bibitem{Hall:1956eh}
H~E Hall and W~F Vinen.
\newblock {The Rotation of Liquid Helium II. II. The Theory of Mutual Friction
  in Uniformly Rotating Helium II}.
\newblock {\em Proc. Roy. Soc. A}, 238(1213):215--234, 1956.

\bibitem{Hohenberg:1977}
P~Hohenberg and B~Halperin.
\newblock {Theory of dynamic critical phenomena}.
\newblock {\em Rev. Mod. Phys.}, 49(3):435--479, 1977.

\bibitem{Imamoglu:1996vwa}
A~{Imamo\u{g}lu} and R~J Ram.
\newblock Quantum dynamics of exciton lasers.
\newblock {\em Phys. Lett. A}, 214(3–4):193 -- 198, 1996.

\bibitem{kadanoff:2000}
L~P Kadanoff.
\newblock {\em Statistical Physics: Statics, Dynamics and Renormalization}.
\newblock World Scientific, Singapore, 2000.

\bibitem{Kasprzak:2006jy}
J~Kasprzak, M~Richard, S~Kundermann, A~Baas, P~Jeambrun, J~M~J Keeling, F~M
  Marchetti, M~H Szyma{\'n}ska, R~Andr{\'e}, J~L Staehli, V~Savona, P~B
  Littlewood, B~Deveaud, and Le~Si Dang.
\newblock {Bose--Einstein condensation of exciton polaritons}.
\newblock {\em Nature}, 443(7110):409--414, 2006.

\bibitem{Keeling:2008hj}
J~Keeling and N~G Berloff.
\newblock {Spontaneous Rotating Vortex Lattices in a Pumped Decaying
  Condensate}.
\newblock {\em Phys. Rev. Lett.}, 100(25):250401, 2008.

\bibitem{Klaers:2010bi}
J~Klaers, J~Schmitt, F~Vewinger, and M~Weitz.
\newblock {Bose--Einstein condensation of photons in an optical microcavity}.
\newblock {\em Nature}, 468(7323):545--548, 2010.

\bibitem{Kneer:1998fma}
B~Kneer, T~Wong, K~Vogel, W~Schleich, and D~Walls.
\newblock {Generic model of an atom laser}.
\newblock {\em Phys. Rev. A}, 58(6):4841--4853, 1998.

\bibitem{kopnin:2001}
N~Kopnin.
\newblock {\em Theory of Nonequilibrium Superconductivity}.
\newblock Oxford University Press, Oxford, 2001.

\bibitem{Krizhanovskii:2009cea}
D~Krizhanovskii, K~Lagoudakis, M~Wouters, B~Pietka, R~Bradley, K~Guda,
  D~Whittaker, M~Skolnick, B~Deveaud-Pl{\'e}dran, M~Richard, R~Andr{\'e}, and
  Le~Dang.
\newblock {Coexisting nonequilibrium condensates with long-range spatial
  coherence in semiconductor microcavities}.
\newblock {\em Phys. Rev. B}, 80(4):045317, 2009.

\bibitem{Kuramoto:1976}
Y~Kuramoto and T~Tsuzuki.
\newblock Persistent propagation of concentration waves in dissipative media
  far from thermal equilibrium.
\newblock {\em Prog. Theor. Phys.}, 55:356, 1976.

\bibitem{Lagoudakis:2008ia}
K~G Lagoudakis, M~Wouters, M~Richard, A~Baas, I~Carusotto, R~Andr{\'e}, Le~Si
  Dang, and B~Deveaud-Pl{\'e}dran.
\newblock {Quantized vortices in an exciton--polariton condensate}.
\newblock {\em Nature Physics}, 4(9):706--710, 2008.

\bibitem{Lega:1994kaa}
J~Lega, J~Moloney, and A~Newell.
\newblock {Swift-Hohenberg Equation for Lasers}.
\newblock {\em Phys. Rev. Lett.}, 73(22):2978--2981, 1994.

\bibitem{Liew:2007kn}
T~Liew, A~Kavokin, and I~Shelykh.
\newblock {Excitation of vortices in semiconductor microcavities}.
\newblock {\em Phys. Rev. B}, 75(24):241301(R), 2007.

\bibitem{Lugiato:1988bta}
L~A Lugiato, C~Oldano, and L~M Narducci.
\newblock {Cooperative frequency locking and stationary spatial structures in
  lasers}.
\newblock {\em J. Opt. Soc. Am. B}, 5(5):879--888, 1988.

\bibitem{Manneville:2004vfa}
P~Manneville.
\newblock Spots and turbulent domains in a model of transitional plane couette
  flow.
\newblock {\em Theoretical and Computational Fluid Dynamics}, 18:169--181,
  2004.
\newblock 10.1007/s00162-004-0142-4.

\bibitem{Matkowsky:2003wba}
B~J Matkowsky and A~A Nepomnyashchy.
\newblock {A complex Swift--Hohenberg equation coupled to the Goldstone mode in
  the nonlinear dynamics of flames}.
\newblock {\em Physica D: Nonlinear Phenomena}, 179(3-4):183, 2003.

\bibitem{Moloney:1995usa}
J~V Moloney and A~C Newell.
\newblock {Universal description of laser dynamics near threshold}.
\newblock {\em Physica D: Nonlinear Phenomena}, 83(4):478--498, 1995.

\bibitem{Penckwitt:2002kda}
A~Penckwitt, R~Ballagh, and C~Gardiner.
\newblock {Nucleation, Growth, and Stabilization of Bose-Einstein Condensate
  Vortex Lattices}.
\newblock {\em Phys. Rev. Lett.}, 89(26):260402, 2002.

\bibitem{PismenBook:1999}
L~M Pismen.
\newblock {\em Vortices in Nonlinear Fields. From Liquid Crystals to
  Superfluids, From Non-Equilibrium Patterns to Cosmic Strings}.
\newblock Oxford University Press, Oxford, 1999.

\bibitem{Roumpos:2010kta}
G~Roumpos, W~H Nitsche, S~H{\"o}fling, A~Forchel, and Y~Yamamoto.
\newblock {Gain-Induced Trapping of Microcavity Exciton Polariton Condensates}.
\newblock {\em Phys. Rev. Lett.}, 104(12):126403, 2010.

\bibitem{Sivashinsky:1977}
G~I Sivashinsky.
\newblock Nonlinear analysis of hydrodynamic instability in laminar flames--i.
  derivation of basic equations.
\newblock {\em Acta Astronaut}, 4:1177--1206, 1977.

\bibitem{Staliunas:1993fk}
K~Staliunas.
\newblock {Laser Ginzburg-Landau equation and laser hydrodynamics}.
\newblock {\em Phys. Rev. A}, 48(2):1573--1581, 1993.

\bibitem{Stringari1996}
S~Stringari.
\newblock {Collective Excitations of a Trapped Bose-Condensed Gas.}
\newblock {\em Phys. Rev. Lett.}, 77(12):2360--2363, 1996.

\bibitem{Stringari1998}
S~Stringari.
\newblock {Dynamics of Bose-Einstein condensed gases in highly deformed traps}.
\newblock {\em Phys. Rev. A}, 58(3):2385--2388, 1998.

\bibitem{Taranenko:1997bt}
V~Taranenko, K~Staliunas, and C~Weiss.
\newblock {Spatial soliton laser: Localized structures in a laser with a
  saturable absorber in a self-imaging resonator}.
\newblock {\em Phys. Rev. A}, 56(2):1582--1591, 1997.

\bibitem{Tosi:2012ika}
G~Tosi, G~Christmann, N~G Berloff, P~Tsotsis, T~Gao, Z~Hatzopoulos, P~G
  Savvidis, and J~J Baumberg.
\newblock {Sculpting oscillators with light within a nonlinear quantum fluid}.
\newblock {\em Nature Physics}, 8:190--194, 2012.

\bibitem{Wertz:2009jka}
E~Wertz, L~Ferrier, D~D Solnyshkov, P~Senellart, D~Bajoni, A~Miard,
  A~Lemaı̂tre, G~Malpuech, and J~Bloch.
\newblock {Spontaneous formation of a polariton condensate in a planar GaAs
  microcavity}.
\newblock {\em Appl. Phys. Lett.}, 95(5):051108--051108--3, 2009.

\bibitem{Wouters:2008vq}
M~Wouters.
\newblock {Excitations and superfluidity in non-equilibrium Bose--Einstein
  condensates of exciton--polaritons}.
\newblock {\em Superlattices and Microstructures}, 43(5--6):524--527, 2008.

\bibitem{Wouters:2007dia}
M~Wouters and I~Carusotto.
\newblock {Excitations in a Nonequilibrium Bose-Einstein Condensate of Exciton
  Polaritons}.
\newblock {\em Phys. Rev. Lett.}, 99(14):140402, 2007.

\bibitem{Wouters:2010ee}
M~Wouters and I~Carusotto.
\newblock {Superfluidity and Critical Velocities in Nonequilibrium
  Bose-Einstein Condensates}.
\newblock {\em Phys. Rev. Lett.}, 105(2):020602, 2010.

\bibitem{Wouters:2010gwb}
M~Wouters, T~Liew, and V~Savona.
\newblock {Energy relaxation in one-dimensional polariton condensates}.
\newblock {\em Phys. Rev. B}, 82(24):245315, 2010.

\end{thebibliography}
%


\printindex
\end{document}